\documentclass[aps,prl,superscriptaddress,twocolumn,longbibliography]{revtex4-1}
\usepackage{graphicx}
\usepackage{amsmath}
\usepackage{amssymb}
\usepackage{amsfonts}
\usepackage{bbold}
\usepackage{layout}
\usepackage{natbib}
\usepackage{dcolumn}
\usepackage{braket}
\usepackage{bm}
\usepackage{color}
\usepackage{verbatim}
\usepackage[colorlinks,allcolors=blue]{hyperref}
\usepackage{url}

\begin{document}

\title{Colossal orbital-Edelstein effect in non-centrosymmetric superconductors}

\author{Luca Chirolli}
\affiliation{Department of Physics, University of California, Berkeley, CA-94720}
\affiliation{NEST, Istituto Nanoscienze-CNR and Scuola Normale Superiore, Piazza San Silvestro 12, I-56127 Pisa, Italy}

\author{Maria Teresa Mercaldo}
\affiliation{Dipartimento di Fisica ``E. R. Caianiello", Universit\`a di Salerno, IT-84084 Fisciano (SA), Italy}

\author{Claudio Guarcello}
\affiliation{Dipartimento di Fisica ``E. R. Caianiello", Universit\`a di Salerno, IT-84084 Fisciano (SA), Italy}

\author{Francesco Giazotto}
\affiliation{NEST, Istituto Nanoscienze-CNR and Scuola Normale Superiore, Piazza San Silvestro 12, I-56127 Pisa, Italy}

\author{Mario Cuoco}
\affiliation{SPIN-CNR, IT-84084 Fisciano (SA), Italy}
\affiliation{Dipartimento di Fisica ``E. R. Caianiello", Universit\`a di Salerno, IT-84084 Fisciano (SA), Italy}

\begin{abstract}
In superconductors that lack inversion symmetry, the flow of supercurrent can induce a non-vanishing magnetization, a phenomenon which is at the heart of non-dissipative magneto-electric effects, also known as Edelstein effects. For electrons carrying spin and orbital moments a question of fundamental relevance deals with the orbital nature of magneto-electric effects in conventional spin-singlet superconductors with Rashba coupling. Remarkably, we find that the supercurrent-induced orbital magnetization is more than one order of magnitude greater than that due to the spin, giving rise to a colossal magneto-electric effect. The induced orbital magnetization is shown to be sign tunable, with the sign change occurring for the Fermi level lying in proximity of avoiding crossing points in the Brillouin zone. In the presence of superconducting phase inhomogeneities, a modulation of the Edelstein signal on the scale of the superconducting coherence length appears, leading to domains with opposite orbital moment orientations. These hallmarks are robust to real-space self-consistent treatment of the superconducting order parameter. The orbital-dominated magneto-electric phenomena, hence, have clear-cut marks for detection both in the bulk and at the edge of the system and are expected to be a general feature of multi-orbital superconductors with inversion symmetry breaking. 
\end{abstract}

\maketitle

{\it Introduction.---} The success in designing spintronic and spinorbitronic devices relies on effects to generate, manipulate and detect spin-polarized currents. The Rashba spin-orbit coupling \cite{Rashba1960,Dresselhaus1955} has a special role in this context, as it allows to directly tune the electron spin orientation through its propagation and, viceversa, to achieve a spin control of the electron trajectories \cite{Manchon2015}, thus giving rise to a large variety of magneto-electric effects. The most notable manifestations are the spin Hall effect \cite{Dyakonov1971, Hirsch1999, Kato2004,Sinova2015} as well as the direct and inverse Edelstein effects \cite{Aronov1989, Edelstein1990, Shen2014}, with a magnetization being induced by an electric current or, conversely, a non-equilibrium magnetization leading to charge current, respectively. 

Recently, it has been recognized that an effect akin to the spin-Rashba (SR) in two-dimensional electron systems (2DES) arises due to the coupling between the atomic orbital angular momentum ${\bf L}$ and the crystal wave-vector ${\bf k}$ \cite{Park2011,Park2012,Kim2013,Mercaldo2020}. For materials having electronic states with non-vanishing ${\bf L}$ close to the Fermi level, an intrinsic crystalline potential or an applied electric field that breaks spatial inversion can yield non-local odd-parity matrix elements among distinct atomic orbitals. The resulting orbital Rashba effect (ORE) \cite{Park2011,Park2012,Kim2013,Mercaldo2020}, in analogy with the spin Rashba effect (SRE), confers chirality to the electronic states at the Fermi level, that acquire a momentum-dependent orbital texture \cite{Park2013,Kim2013,Park2012b,Kim2012,Mercaldo2020}. Interestingly, materials that have negligible atomic spin-orbit interaction and that would be ruled out from spin-orbitronics applications, now emerge as a novel platform were orbitronic effects can become relevant \cite{Ding2022}.

Orbital analog phenomena of the spin Hall and spin Edelstein effects have been proposed, namely orbital Hall \cite{Zhang2005, Tanaka2008, Kontani2008} and orbital Edelstein effects \cite{Levitov1985,Zhong2016,Salemi2019,Johansson2021}. The former arises from the electrons’ orbital motion \cite{VanderbiltResta2005} and in 2D systems only affects the out-of-plane magnetization. The latter has its origin in the atomic orbital content of the wavefunction and leads to a current-induced in-plane orbital magnetization. 
Brought into the context of non-dissipative superconducting magnetoelectric effects, so far the focus has been mostly on the spin degrees of freedom for spin-Rashba type superconductors \cite{Edelstein1995, Edelstein2005, Yip2014, Konschelle2015, Tokatly2017,He2019,He2021} including the generalization to all gyrotropic crystal point groups \cite{He2020}.  However, since electrons carry both spin and orbital moments, assessing the role of electrons' orbital moment in setting out the magneto-electric effects is of fundamental relevance. Furthermore, the emergence of materials that show strong orbital-Rashba coupling and the possibility to induce superconductivity by the proximity effect promote the novel field of superconducting orbitronics, opening unprecedented scenarios and applications.

In this manuscript, we study the emergence of magneto-electric effects induced by a supercurrent flow in 2D non-centrosymmetric superconductors featuring an orbital Rashba (OR) coupling. We show that the supercurrent-induced orbital magnetization can exhibit an extraordinary large amplitude and a peculiar sign tunability. The orbital moment is more than one order of magnitude greater than the spin one, for nominally equal coupling strength. The orbital Edelstein effect becomes particularly enhanced for the Fermi level lying in proximity of avoiding crossing points in the band structure induced by the orbital Rashba coupling, where it also features a sign change. The orientation of the induced orbital magnetization can be suitably tuned by sweeping the Fermi level across the avoided crossing, providing a knob to master the response. The spatial dependence of the signal reflects inhomogeneities of current flow on the scale of the superconducting coherence length, featuring sign changes and signal enhancement in proximity of superconducting phase defects. This outcome provides a route to spatially resolve supercurrent amplitude variations and to devise domains with opposite magnetization without applied magnetic fields. 


\begin{figure}[t]
\includegraphics[width=0.45\textwidth]{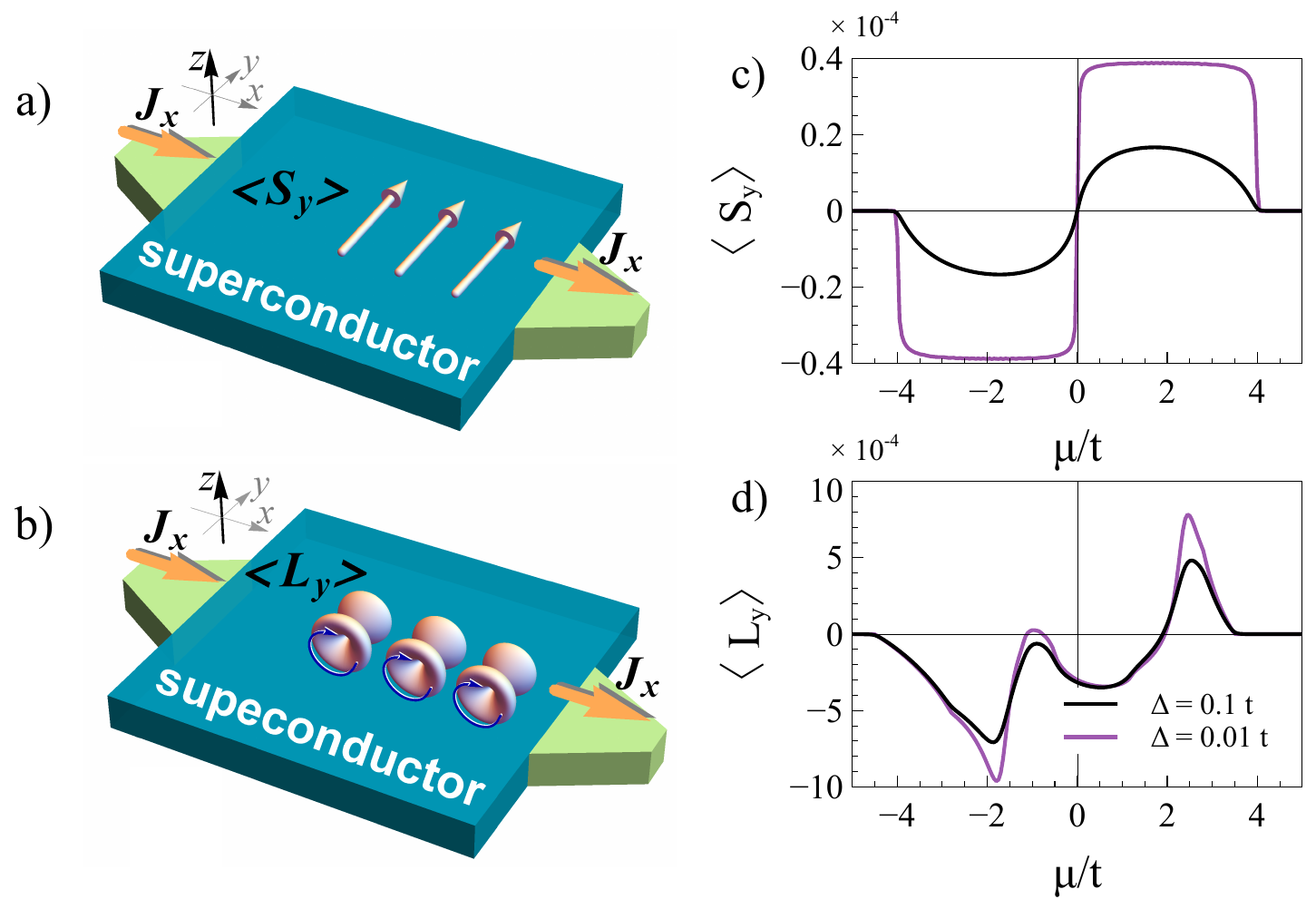}
\caption{Schematic setup showing the Edelstein effect in (a) spin- and (b) orbital-Rashba 2D systems with a supercurrent $J_x$ flowing along the $x$-axis and the corresponding induced magnetization in the transverse $y$-direction. Induced (c) spin- and (d) orbital-polarization integrated in the whole Brillouin zone for a 2D Rashba system as a function of the chemical potential for {two values of the superconducting gap, $\Delta = 0.1 t, 0.01 t$}, $\alpha=0.1 t$ and $q=0.02$ in unit of the inverse atomic distance. In (d) the employed parameters are $t'=0.4 t$ and $\delta_z=0.5 t$.
\label{Fig1}}
\end{figure}

{\it Model.---} We consider a 2D electronic system that explicitly includes either the spin or orbital Rashba couplings, specified by a unique coupling $\alpha$. For the OR case, the minimal description is based on three bands arising from atomic orbitals spanning the $L=1$ angular momentum subspace, such as $p$ orbitals or $d_a$ orbitals with $a=(yz,xz,xy)$. For definiteness we refer to $d$ orbitals localized at the site of a square lattice. The breaking of mirror symmetry, due to intrinsic crystalline potential or externally applied electric field, sets out a polar axis ($z$), resulting in a $C_{4v}$ point group symmetry, and an orbital Rashba interaction that couples the atomic angular momentum ${\bf L}$ with the crystal wave-vector ${\bf k}$. The Hamiltonian in momentum space is written as
\begin{equation}\label{Eq:h0}
h_0({\bf k})=\sum_a\epsilon_{a}({\bf k})\mathbb{P}_a+\alpha \left( {\boldsymbol{\gamma}_{\bf k}}\wedge \hat{\bf L} \right) \cdot \hat{z}-\mu,
\end{equation}
where $\epsilon_a({\bf k})=-2t^x_a\cos(k_x)-2t^y_a\cos(k_y)-\delta_a$ are dispersion relations resulting from symmetry-allowed nearest-neighbor hopping amplitudes, with $t_{yz}^y=t_{xz}^x=t_{xy}^x=t_{xy}^y=t$, $t_{yz}^x=t_{xz}^y=t'$, and $\mathbb{P}_a=(\hat{L}^2-2\hat{L}_a^2)/2$. At the $\Gamma$ point a degeneracy occurs between, say, the $xz$ and the $yz$ orbitals, and we set $\delta_{xz}=\delta_{yz}=0$. In turn, the orbital $xy$ can show a finite crystal field splitting $\delta_{xy}\equiv \delta_z$. The three bands are coupled at finite momentum by the OR interaction with coupling constant $\alpha$ and involving the components of the angular momentum $[\hat{L}_k]_{lm}=i \epsilon_{klm}$, with $\epsilon_{klm}$ the Levi-Civita tensor. The OR coupling is specified by a vector ${\boldsymbol{\gamma}({\bf k})}=(\sin(k_x),\sin(k_y),0)$ in momentum space that sets out the chirality of the orbital texture. We then assume a finite chemical potential and a purely local ($s-$wave) attractive interaction that opens a superconducting gap of strength $\Delta$ on the Fermi surface.

{\it Orbital Edelstein effect.---} 
The Edelstein effect in SR coupled superconductors manifests as a finite spin polarization in response to an applied supercurrent (Fig.~\ref{Fig1}(a)). The wedge product of the polar vector associated to the mirror symmetry breaking direction, ${\bf z}$, and the supercurrent ${\bf J}$, allows the construction of an axial vector 
\begin{equation}\label{Eq:LinearResp}
{\bf M} \sim {\bf z}\wedge {\bf J},
\end{equation}
with a spin magnetization orthogonal to the applied bias current (Fig.~\ref{Fig1}(a)). Analogously, in OR coupled superconductors an orbital Edelstein effect is expected, with a finite orbital magnetization ${\bf M}=\langle{\bf L}\rangle$ (in unit of the Bohr magneton $\mu_B$) orthogonal to the polar axis ${\bf z}$ and supercurrent ${\bf J}$, as Fig. \ref{Fig1}(b). We point out that for the examined 2D system an in-plane component of the orbital magnetization is not expected to take any contribution from a non-trivial out-of-plane Berry curvature.

In the presence of a current bias the order parameter acquires a position-dependent phase, $\hat\Delta({\bf r})=\hat{\Delta} e^{i{\bf q}\cdot{\bf r}}$ for a finite ${\bf q}$ sustained by the bias current (see Supplementary Material \cite{Note1}). 
The position dependence of the gap can be mapped via a gauge transformation in a momentum shift ${\bf q}/2$ of opposite sign for particle and holes. For weak values of ${\bf q}$ the magnetization can be written in linear response theory as
\begin{equation}\label{Eq:Lchi}
\langle \hat{L}_\mu\rangle=-\frac{1}{2}\chi_{\mu\nu}q_\nu,
\end{equation}
with $\chi_{\mu\nu}$ a static susceptibility between the angular momentum $\hat{L}_\mu$ and the velocity operator $\hat{v}_\mu=\partial h_0/\partial k_\mu$. The latter is composed by a normal and an anomalous term and acquires no structure in particle-hole space.  The presence of a finite superconducting gap in the spectrum allows us to write the static susceptibility by energy correction at second-order perturbation theory and is determined by the matrix elements $v_x^{n,n'}L_y^{n'n}\equiv\langle\phi_{n,{\bf k}}|\hat{v}_x|\phi_{n',{\bf k}}\rangle\langle\phi_{n',{\bf k}}|\hat{L}_y|\phi_{n,{\bf k}}\rangle$, with $|\phi_{n,{\bf k}}\rangle$ eigenstates of $h_0({\bf k})$, Eq.~(\ref{Eq:h0}). 

\begin{figure}[t]
\includegraphics[width=0.45\textwidth]{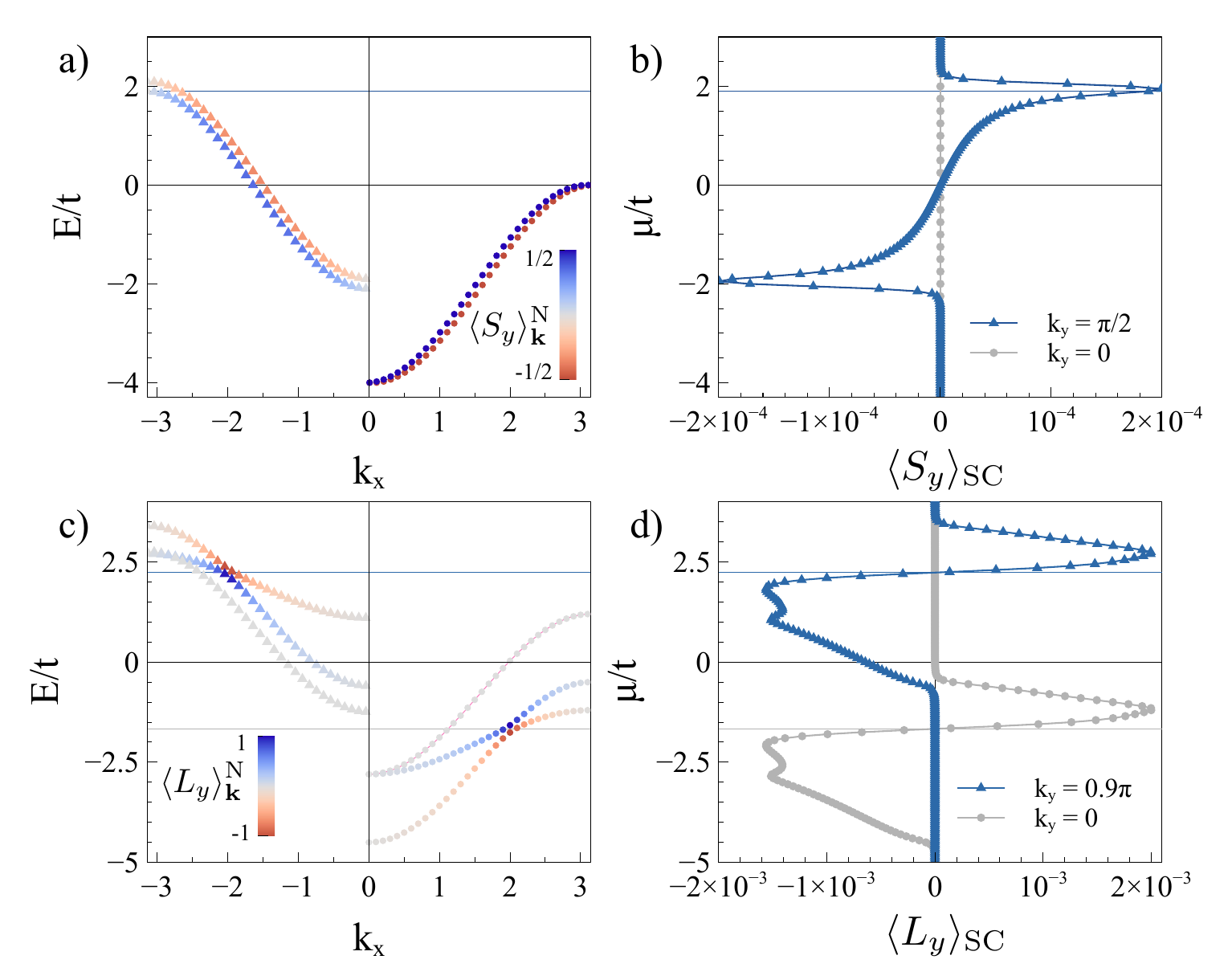}
\caption{(a) Spin-split Rashba bands for $k_y=0$ (for $k_x>0$) and $k_y=\pi/2$ (for $k_x<0$). The color map encodes amplitude and sign of the $k$-resolved spin polarization in the normal state. (b) Energy-resolved spin polarization for the spin-Rashba model in the superconducting state, for the two cuts in the Brillouin zone shown in (a). (c) Cuts of the orbital-resolved band structure at $k_y=0$ (for $k_x>0$) and at $k_y=0.9 \pi$ (for $k_y<0$), respectively. The color map indicates the amplitude and sign of the $k$-resolved angular momentum polarization in the normal state. (d) Induced orbital moment versus energy in the superconducting state for the two cuts reported in (c). Employed parameters as in Fig.~\ref{Fig1} with $\Delta=0.1 t$.\label{Fig2}}
\end{figure}

The comparison of the outcome of the total spin and orbital moments for the SR and OR cases is reported in Fig.~\ref{Fig1}(c),(d). Two striking features can be immediately observed: i) the orbital moment is much larger than the spin moment, with a maximum up to {25 (60) times the spin one for $\Delta = 0.01 t$ ($\Delta=0.1 t$)} and for equal coupling constant, ii) the signal changes sign at a given chemical potential. As we will discuss in detail, these features have their origin in a band structure featuring avoided crossings  (Fig.~\ref{Fig2} (c) and (d)).  The latter can occur in generic points in the Brillouin zone, in that orbitals, differently from spin,   are not forced to be degenerate at high symmetry point by Kramers' theorem. The combination of orbital symmetry and the crystal point group of the material may allow a crystal field splitting and a difference in effective masses to appear, resulting in avoided crossings and orbital tunable magneto-electric effects not only in semi-metallic but also in metallic states. In contrast, the sign change of the spin moment for the SRE is due to the sublattice symmetry of the tight-binding model at specific points in parameter space and manifests at $\mu=0$.

{\it The role of avoided crossings.---}
A simple way to understand the difference in spin and orbital response is to consider 1D cuts of the band structure by fixing the transverse momentum $k_y$. In particular, for $k_y=0,\pi$ one band in the OR case completely decouples and the only relevant matrix element reads \cite{Note1}
\begin{equation}\label{Eq:MatElky0}
v_x^{+,-}L^{-,+}_y=\alpha(\epsilon_1-\epsilon_2)\frac{(\epsilon_1-\epsilon_2)\partial_x{\gamma_x}-{\gamma_x}(v^1_x-v^2_x)}{4\alpha^2{\gamma_x^2}+(\epsilon_1-\epsilon_2)^2},
\end{equation}
with $\partial_x\equiv\partial/\partial k_x$, $v_{1(2)}=\partial_x \epsilon_{1(2)}(k_x)$, and $\epsilon_{1(2)}=\epsilon_{yz(xz)}(k_x)$. In Fig.~\ref{Fig2}(a) we show the bands at $k_y=0 (\pi/2)$ for positive (negative) $k_x$ of the SR model for $\alpha=0.1 t$ and in Fig.~\ref{Fig2}(b) the induced spin magnetization evaluated for $q=0.02$ (in units of the inverse atomic lattice constant), for which the linear response condition Eq.~(\ref{Eq:LinearResp}) is satisfied. Clearly, in the case of SR, the unperturbed spin-degenerate bands satisfy $\epsilon_1=\epsilon_2$ and the Edelstein susceptibility is exactly zero (Fig. \ref{Fig2}(b) circular marks).  At finite momentum $k_y$ the unperturbed bands admix and the susceptibility in the SR case becomes finite (Fig. \ref{Fig2}(b) triangular marks), with a response peaked at the band splitting nearby the $\Gamma$ point. 

The orbital Edelstein magnetization is more interesting. In Fig.~\ref{Fig2}(c) we show the bands at $k_y=0 (0.9\pi)$ for positive (negative) $k_x$ of the OR model Eq.~(\ref{Eq:h0}) for $t'=0.4 t$ and $\alpha=0.1 t$. A light effective mass band, originating from the $d_{xy}$ orbital, detaches and shifts down in energy with respect to  the other two heavy bands, that remain degenerate at the $\Gamma$ point. As a result the light and heavy bands cross at finite momentum and an avoided crossing is generated for finite $\alpha$.  In Fig.~\ref{Fig2}(d) the induced orbital moment for $q=0.02$ is finite for every $k_y$  and a sign change appears that is pinned to the avoided crossing. This is understood at $k_y=0$ by inspection of Eq.~(\ref{Eq:MatElky0}). At the avoided crossing, the energy difference between the unperturbed bands $\epsilon_1-\epsilon_2$ changes sign and so does the matrix element Eq.~(\ref{Eq:MatElky0}). Alternatively,  the velocities of the Rashba-split bands interchange at the avoided crossing. Their sum, weighted by the angular momentum expectation value $\langle L_y\rangle^{\rm N}_{\bf k}$, of opposite sign for the two bands and maximal at the avoided crossings (see color code in Fig.~\ref{Fig2}(c)), gives rise to a sign change of the signal and to peaks in its proximity. 

\begin{figure}[t!]
\includegraphics[width=0.45\textwidth]{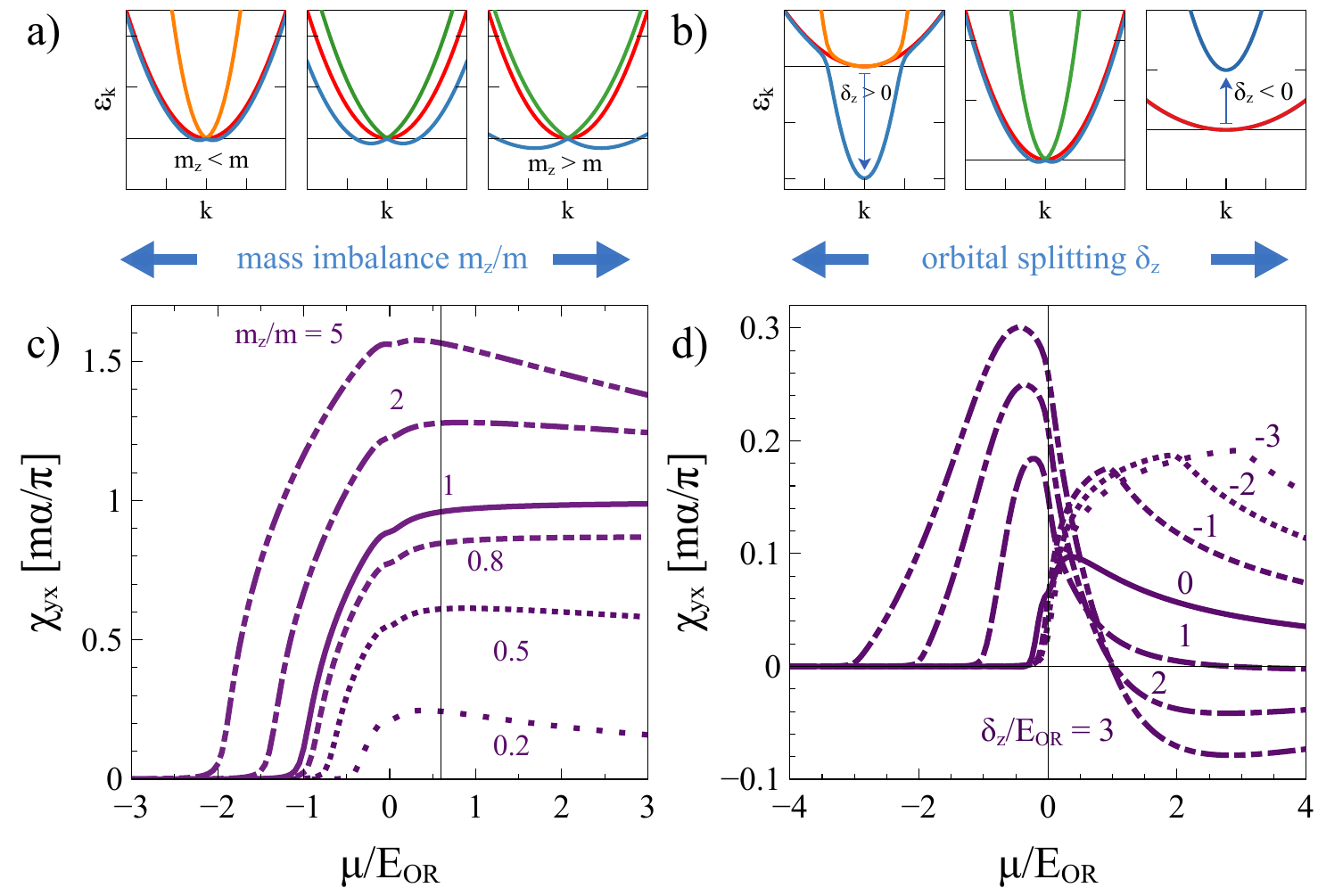}
\caption{Evolution of the bands around the $\Gamma$ point in the $C_{\infty v}$ symmetric case by varying a) the orbital mass imbalance $m_z/m$ for $\delta_z=0$ and b) the crystal field splitting $\delta_z$ for $m_z=5m$. Edelstein susceptibility versus the chemical potential $\mu$ in units of $E_{\rm OR}=m\alpha^2/2$ by varying c) the orbital mass ratio $m_z/m$ for $\delta_z=0$ and d) the crystal field splitting $\delta_z/E_{\rm OR}$ for $m_z=5m$. 
\label{Fig3}}
\end{figure}

{\it Orbital splitting and mass imbalance.---}
To further stress the band structure origin of the Edelstein effect and differences between SRE and ORE, we simplify the analysis by expanding the Hamiltonian around the $\Gamma$ point, where the OR term acquires a $C_{\infty v}$ symmetry, and extend the latter to the effective masses.  This way, the avoided crossing in Fig.~\ref{Fig2} are now at the same energy for all $k_x$ and $k_y$ and the strong dependence of the Edelstein response in effective masses and crystal field splitting becomes manifest.

\begin{figure}[t!]
\includegraphics[width=0.45\textwidth]{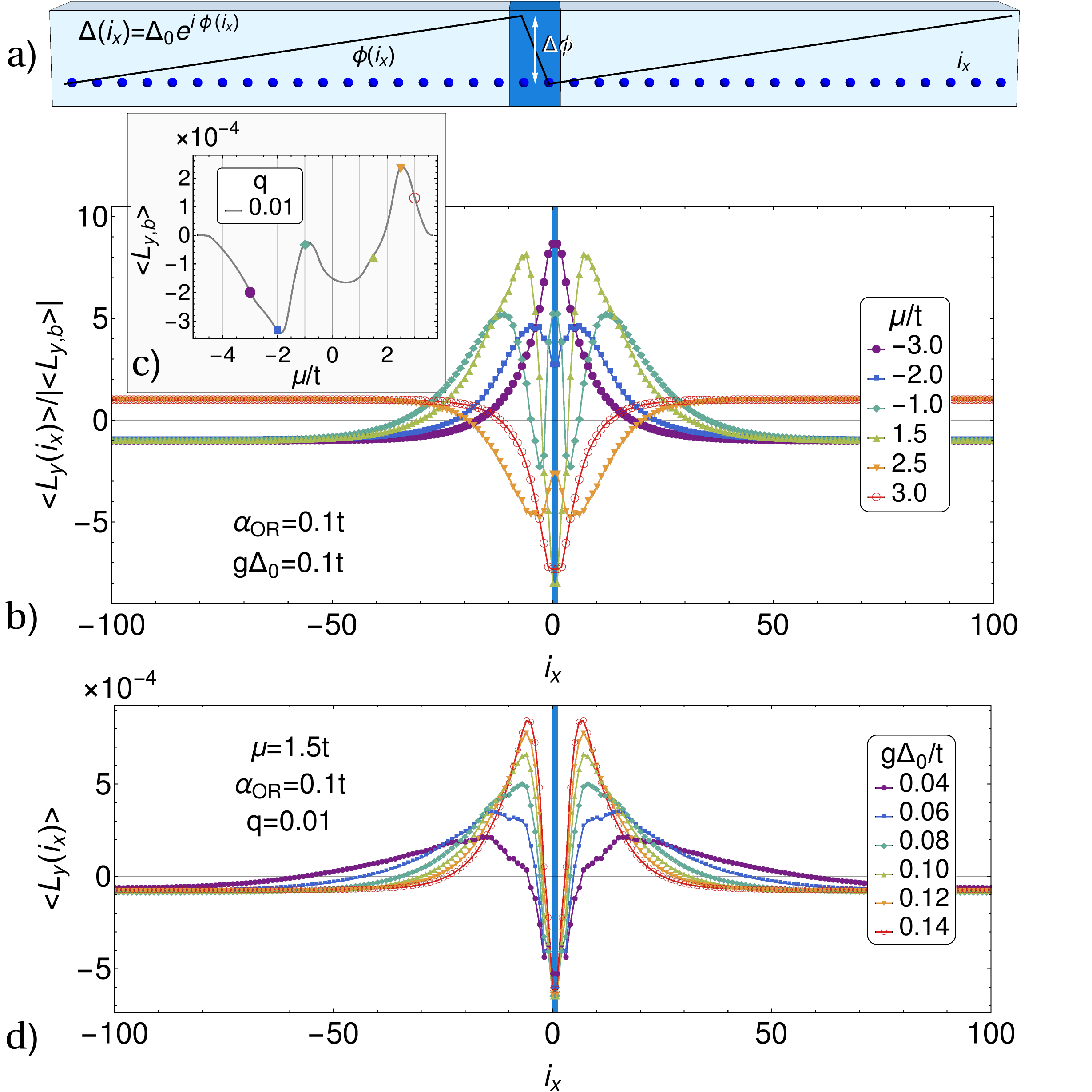}
\caption{(a) Sketch of the superconducting phase across the superconductor along a given direction in real space, with a change of the phase gradient by $\Delta \phi$ around the position $i_x=0$. (b) Real space dependence of the orbital polarization renormalized to the bulk value $\langle L_{y,b} \rangle$ reported in (c) for several values of the chemical potential. In proximity of the domain boundary ($i_x=0$) the orbital moment has an enhancement of the amplitude of about 10 times with respect to the bulk, and a sign change occurring at a distance of about the superconducting coherence length. (d) Spatially resolved orbital moment for a representative value of $\mu$ and $\alpha$ for different values of the superconducting gap $\Delta_0$. The increase of the gap amplitude induces a sign change of the orbital polarization at a shorter distance from the phase domain boundary thus scaling with the coherence length amplitude. The employed parameters are $t'=0.4 t$, $\delta_z=0.5 t$, while the hopping amplitude at the domain boundary is $t_b=0.75 t$. 
\label{Fig4}}
\end{figure}

In Fig.~\ref{Fig3}(a) we show the evolution of the bands by varying the ratio $m_z/m$. For $m_z/m\neq 1$ two of the three bands tend to collapse on top of the other, so that their relative contribution vanishes. The susceptibility, shown in Fig.~\ref{Fig3}(c) is dominated by the residual contribution, that is proportional to the difference in the Fermi momenta and for $m_z/m>1$ the signal increases. In turn, the Edelstein response for the SRE case around the $\Gamma$ point is featureless, showing no sign change and a constant susceptibility \cite{He2019} resembling the $m=m_z$ line in Fig.~\ref{Fig3} (c).  In Fig.~\ref{Fig3}(b) we show the evolution of the bands by varying the value of the crystal field splitting $\delta_z$.  Choosing $m=5 m_z$ the band edge of the light mass band shifts down (up) in energy for $\delta_z>0 ~(\delta_z <0)$. An avoided crossing is then produced for $\delta_z>0$, as in the case illustrated in Fig.~\ref{Fig2}(c), that manifests in a sign change in the susceptibility, as shown in Fig.~\ref{Fig3}(d).

\begin{figure}[tb]
\includegraphics[width=0.45\textwidth]{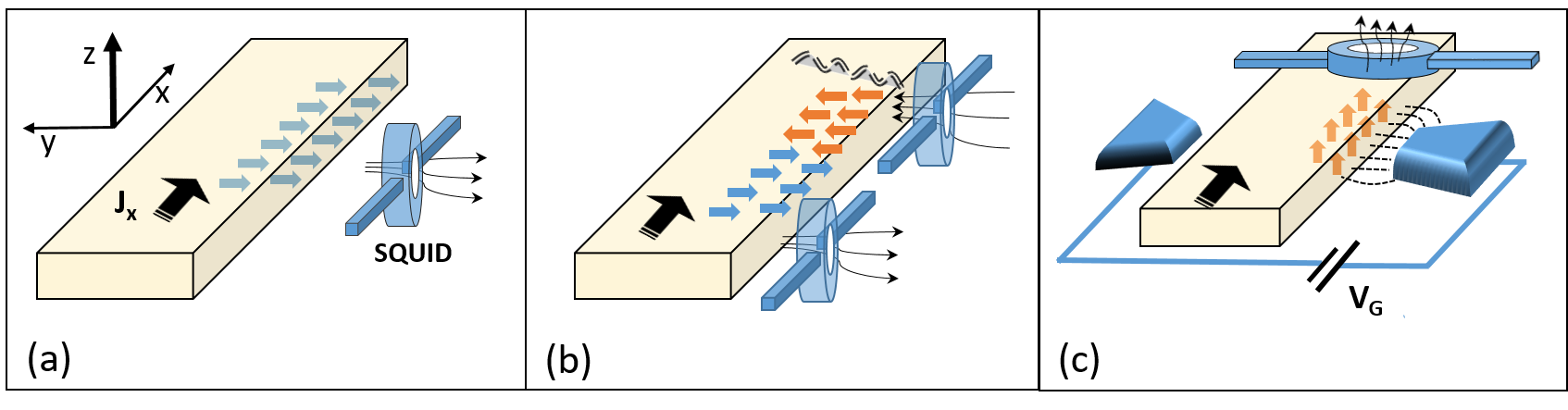}
\caption{Sketch of an experimental setup for the detection of the orbital moments generated by the supercurrent flow. For systems in which $z$ is the symmetry breaking direction,  a lateral SQUID can measure the orbital moment in: (a) uniform magnetization, (b) in presence of small phase inhomogeneities, that generates a  reorientation of the orbital moments in real space. (c) The application of an electric field with lateral split gates leads to an out-of-plane orbital polarization and in turn a magnetic flux that is oriented perpendicular to the surface of the superconductor.  
\label{Fig5}}
\end{figure}

{\it Spatial-dependent orbital Edelstein effects. ---} The amplitude and sign of the ORE has also distinctive marks when considering its spatial dependence in the case of inhomogeneous superconducting phase profile. This aspect is highly peculiar of the superconducting state. We have solved the OR model Eq.~(1) for a 2D system with size $L_x \times L_y$, $L_x=200$, in unit of the inter-atomic distance, and assuming translational invariance for $L_y$. The supercurrent profile includes a phase gradient in real space along a given direction and a phase defect in the spatial profile of the supercurrent (Fig. \ref{Fig4}(a)). The latter represents a domain boundary with a weak modification of the phase gradient by $\Delta \phi$ across the domain boundary at a given position.  The presence of a weak phase gradient change leads to two striking features. We find a general enhancement of the induced orbital moment (Fig. \ref{Fig4}(b))  compared to the value in the bulk (Fig. \ref{Fig4}(c)), that further increases the colossal character of the signal. Moreover, the orbital Edelstein effect exhibits a sign variation of the orbital moment in real space thus yielding domains with opposite orientation. The spatial extension of the domain with an amplification of the orbital polarization is of the order of the superconducting coherence length $\xi_{SC}$. We verify that for a ballistic superconductor, for which $\xi_{SC}\sim v_F/\Delta$ with $v_F$ the Fermi velocity, an increase of the gap amplitude leads to a reduction of the size of the region with enhanced orbital polarization (Fig.~\ref{Fig4}(d)). This result fully pertains the superconducting state and is not expected to have a counterpart in the normal state, where the length scales associated are much shorter, especially in the case of a metal. Interestingly, the spatial variation of the Edelstein response appearing in Fig.~\ref{Fig4} goes beyond the linear response theory result, although still remaining of long wavelength character. These results show how the Edelstein response can be generally exploited to monitor and detect phase disorder patterns in a non-centrosymmetric superconductor.
We have also verified that the effect is robust when allowing for a self-consistent solution of the superconducting order parameter in real space in the presence of supercurrent and phase spatial defects. We report weak variation of the amplitude, with an upper bound of about ten percent, for an orbital Rashba coupling that is smaller than the kinetic energy \cite{Note1}.

{\it Discussion.---} 
The amplitude of the orbital contribution is extraordinary large compared to the spin one for nominally equal coupling constant $\alpha$. The magnitude of the OR coupling constant is typically much larger \cite{Go2017,Park2020a,Go2021,Sunko2017,Unzelmann2020} than the SR one, as it relies on a  momentum-dependent finite overlap between Wannier orbitals enabled by the broken mirror symmetry. The resulting colossal orbital Edelstein response shows wide amplitude and sign tunability through a gate \cite{DeSimoni2018}. Ideal platforms are represented by 2D superconductors realized at LAO/STO (LaAlO$_{3}$/SrTiO$_{3}$) interfaces \cite{Reyren2007,Caviglia2008}, as well as ultrathin elemental superconductors and hybrid implementations through proximitized superconductivity in copper oxides \cite{Ding2022}. 

The robustness of the effect is assessed by considering the self-consistent solution of the gap amplitude and phase. The outcome also suggests that the effect has 
a weak detrimental feedback on the superconducting state and, although small in amplitude, might be detectable in observables that are directly related to the strength of the superconducting state. Detection of the Edelstein effect points to magnetoscopy to probe the spatial distribution of the local dipolar magnetization field, for an in-plane field (Fig.~\ref{Fig5} (a)) or an out-of-plane field generated via a side gate (Fig.~\ref{Fig5} (c)). A SQUID-based technique can also directly probe sign change occurring around phase defects or similar inhomogeneities (Fig.~\ref{Fig5} (b)). The size of the induced local orbital moment can reach about $10^{-4} \mu_{\text{B}}$, and even higher by varying the OR coupling, corresponding to hundred Gauss in close proximity of the 2DES.

{\it Acknowledgments.---}
This project has received funding from the European Union’s Horizon 2020 research and innovation programme under the Marie Sklodowska-Curie grant agreement No 841894 (TOPOCIRCUS). M.C., M.T.M and F.G. acknowledge support by the EU’s Horizon 2020 research and innovation program under Grant Agreement nr. 964398 (SUPERGATE). F.G. acknowledges the European Research Council under Grant Agreement No. 899315-TERASEC, and  the  EU’s  Horizon 2020 research and innovation program under Grant Agreement No. 800923 (SUPERTED) for partial financial support.

\end{document}


\title{Supplemental Material: Colossal orbital-Edelstein effect in non-centrosymmetric superconductors}

\maketitle
\section{Edelstein susceptibility}

We associate fermionic annihilation operators $\hat{d}_{a,s}({\bf r})$ to $d_a$ orbitals at position ${\bf r}$ with spin $s$ and describe $s$-wave superconductivity by assuming a purely local attractive interaction of strength $-g<0$. The latter leads to pairing of electrons with same orbital character in the spin-singlet channel and to the opening of a superconducting uniform gap $\Delta=-g\langle \hat{d}_{a,\downarrow}({\bf r})\hat{d}_{a,\uparrow}({\bf r})\rangle$ on the Fermi surface.

BCS mean-field superconductivity is described by introducing the Nambu basis, $\Psi^\dag({\bf r})=(\hat{\bf d}^\dag_{\uparrow}({\bf r}),\hat{\bf d}^T_{\downarrow}({\bf r}))$, with $\hat{\bf d}_{s}({\bf r})$ a column vector of annihilation Fermionic operators $\hat{d}_{a,s}$ associated to the different orbitals. The mean-field Hamiltonian is then written as $H=\int d{\bf r}\Psi^\dag({\bf r}){\cal H}({\bf r})\Psi({\bf r})$, with ${\cal H}({\bf r})$ the Bogoliubov deGennes Hamiltonian featuring an explicit position dependence via the phase of the superconducting gap $\Delta({\bf r})=\Delta e^{i{\bf q}\cdot{\bf r}}$, with ${\bf q}$ sustained by a finite bias current. Performing a gauge transformation on particle and hole states, $\Psi({\bf r})\to e^{i{\bf q}\cdot{\bf r}\tau_z/2}\Psi({\bf r})$, we can map the position dependence of the gap in a momentum shift of opposite sign for particle and holes, so that the BdG Hamiltonian in momentum space takes the form
\begin{equation}\label{Eq:HbdgQ}
{\cal H}({\bf k},{\bf q})=\left(\begin{array}{cc}
h_0({\bf k}+{\bf q}/2) & \hat{\Delta}({\bf q})\\
\hat{\Delta}^\dag({\bf q}) & -h^*_0(-{\bf k}+{\bf q}/2)
\end{array}\right),
\end{equation} 
where $\hat{\Delta}({\bf q})$ is a gap matrix proportional to the identity and we neglect the momentum dependence of the gap amplitude. 

Following the same approach employed for the spin Rashba coupled superconductor, in the OR case the magnetization in response to an applied supercurrent is given by $\langle \hat{L}_\mu\rangle=T\sum_{i\omega_n,{\bf k}}{\rm Tr}[G_{i\omega_n}({\bf k},{\bf q})\hat{L}_\mu]$, where the Green's function in the Nambu space is given by $-G^{-1}_{i\omega_n}({\bf k},{\bf q})=-i\omega_n+{\cal H}({\bf k},{\bf q})$. Expansion of the full Green's function at first order in ${\bf q}$ yields the expression Eq.~(3) of the main text in terms of a static susceptibility
\begin{equation}
\chi_{\mu\nu}=-T\sum_{i\omega_n,{\bf k}}{\rm Tr}\left[G_{i\omega_n}({\bf k})\hat{L}_\mu G_{i\omega_n}({\bf k}) \hat{v}_\nu({\bf k})\right],
\end{equation}
with the velocity operator is explicitly given by
\begin{eqnarray}\label{Eq:v}
\hat{v}_\mu({\bf k})&=&\sum_a\frac{\partial \epsilon_a({\bf k})}{\partial k_\mu}\mathbb{P}_a+\alpha \hat{\bf L}\wedge \hat{z}\cdot\frac{\partial {\bf g}({\bf k})}{\partial k_\mu}.
\end{eqnarray}
Assuming to apply the current bias along the $x$ direction, the susceptibility can be written as
\begin{eqnarray}\label{Eq:ChiUniformGap}
\chi_{yx}&=&2\sum_{{\bf k},n> n'}\frac{1}{E_{n,k}+E_{n',k}}\left(1-\frac{\epsilon_{n,k}\epsilon_{n',k}+\Delta^2}{E_{n,k}E_{n',k}}\right)\nonumber\\
&\times&{\rm Re}\langle\phi_{n,{\bf k}}|\hat{v}_x|\phi_{n',{\bf k}}\rangle\langle\phi_{n',{\bf k}}|\hat{L}_y|\phi_{n,{\bf k}}\rangle.
\end{eqnarray}
This expression is easily adapted to describe the Edelstein response in the spin-Rashba case by replacing the dispersion $\epsilon_a({\bf k})$ with degenerate bands, the angular momentum $\hat{L}_\mu$ with spin Pauli matrices, and an overall factor 1/2 to avoid double counting spin-degeneracy. This way, a direct comparison between the two models can be made.

\begin{figure}[t]
\includegraphics[width=0.45\textwidth]{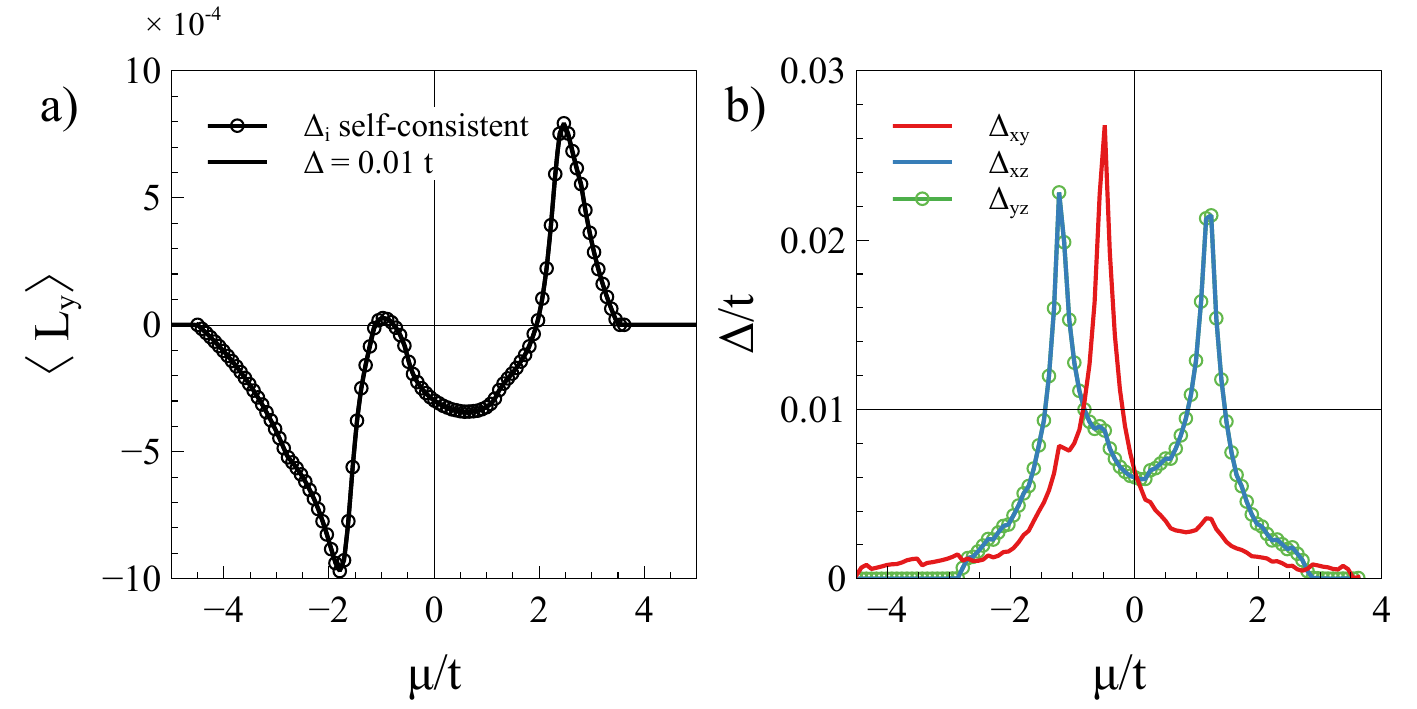}
\caption{(a) Value of the orbital moment $\langle L_y\rangle$ for a homogeneous system obtained for equal gaps $\Delta = 0.01 t$ and for the values shown in (b) that have been obtained self-consistently choosing $g=0.8 t$.
\label{Fig0sm}}
\end{figure}

In the $C_{v\infty}$ case around the $\Gamma$ point the eigenvalues of the band Hamiltonian $h_0$ have the general expression $\epsilon_{0,k}=\epsilon_k$ and $\epsilon_{\pm,k}=\epsilon_k+\delta\epsilon_{\pm,k}$,
with $\delta\epsilon_{\pm,k}=(\epsilon^z_k-\epsilon_k\pm\sqrt{4k^2 \alpha^2+(\epsilon^z_k-\epsilon_k)^2})/2$, $\epsilon^z_k=k^2/2m_z-\mu-\delta_z$, and $\epsilon_k=k^2/2m-\mu$. The matrix elements that enter the susceptibility are then given by
\begin{eqnarray}\label{Eq:AvMatEl}
\left\langle v_x^{0,\pm}L^{\pm,0}_y\right\rangle_\phi&=&\pm\frac{\alpha\delta\epsilon_\pm}{2(\epsilon_k^+-\epsilon_k^-)},\\
\left\langle v_x^{+,-}L^{-,+}_y\right\rangle_\phi&=&\frac{\alpha(\epsilon_k-\epsilon^z_k)}{(\epsilon^+_k-\epsilon^-_k)^2}\left(\epsilon_k-\epsilon^z_k+\frac{\alpha k^2 }{m_z}-\frac{\alpha k^2 }{m}\right),\nonumber\\
\end{eqnarray} 
where $\langle\ldots\rangle_\phi$ denotes average over the azimuthal angle $\phi=\tan^{-1}(k_y/k_x)$. We see that $\left\langle v_x^{+,-}L^{-,+}_y\right\rangle_\phi$ essentially agrees with Eq.~(4) of the main text. 

\begin{figure*}[tb]
\includegraphics[width=0.95\textwidth]{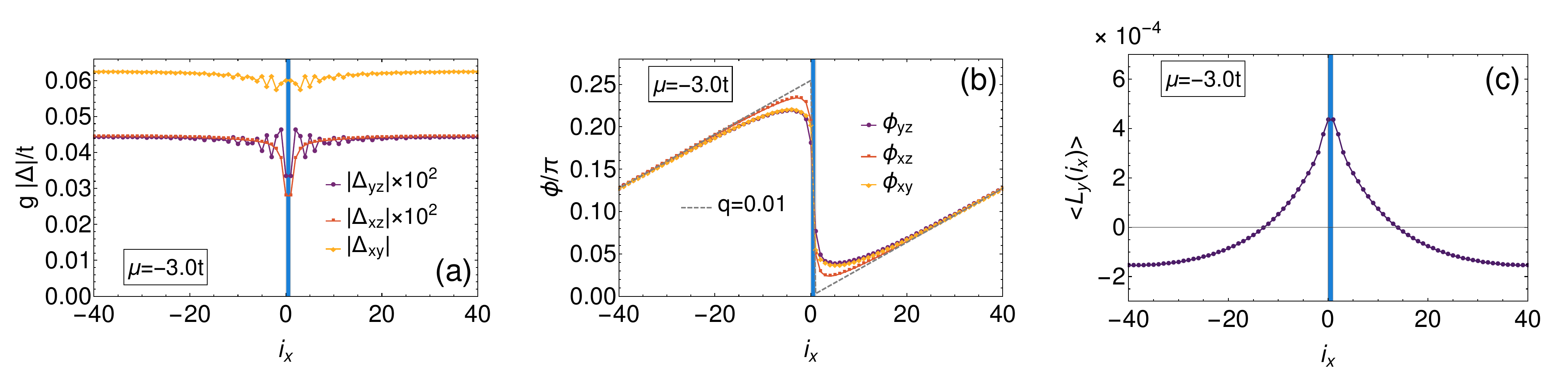}
\includegraphics[width=0.95\textwidth]{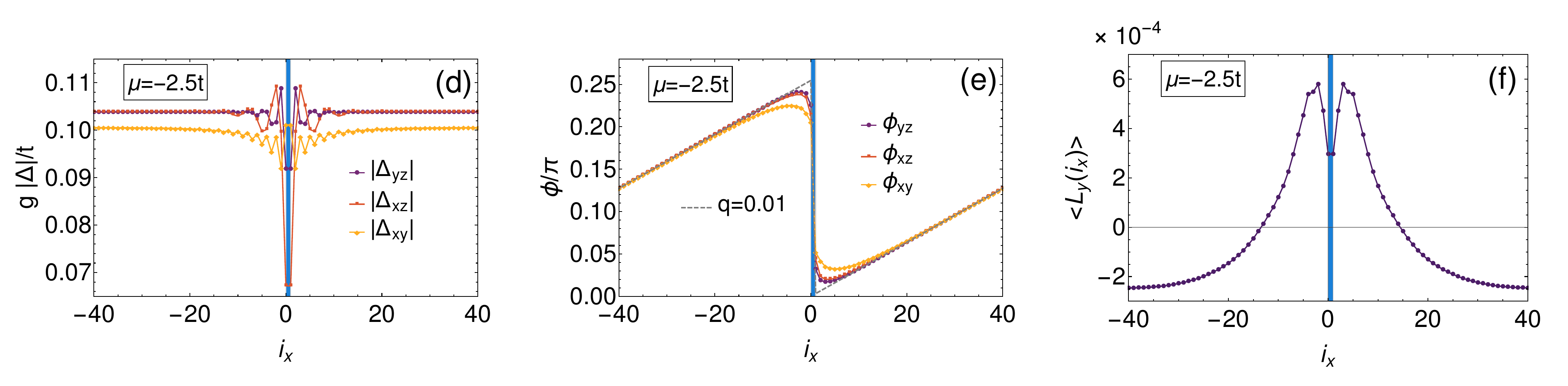}
\includegraphics[width=0.95\textwidth]{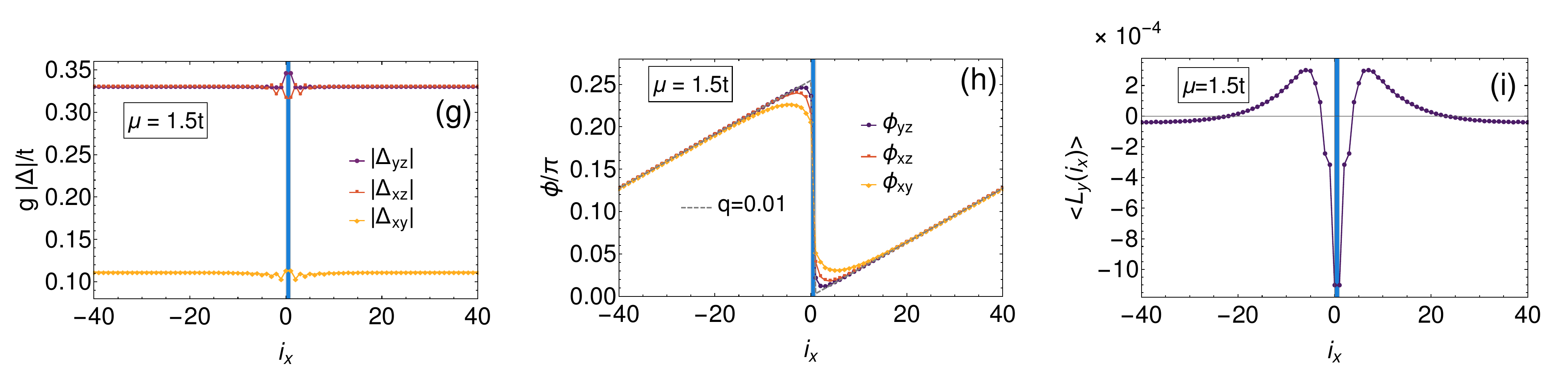}
\caption{(a) Spatial profile of the amplitude of the self-consistent orbital-dependent superconducting order parameters for each orbital character, i.e. $\Delta_{xz}$, $\Delta_{yz}$, $\Delta_{xy}$, in proximity of a region with non-uniform phase gradient around the position $i_x=0$ as reported in panel (b). (c) Real space dependence of the orbital moment $\langle L_{y} \rangle$ for $\mu=-3.0 t$. We observe that the sign change and the amplitude of the orbital moments are not affected by the spatial variation of the order parameters as due to the self-consistent procedure. (d)-(f) As in panels (a)-(c) but for a chemical potential $\mu=-2.5 t$ and (g)-(i) for $\mu=1.5 t$ . For this electron filling the amplitude of the superconducting order parameter also exhibits a reconstruction close to the phase defect. This spatial variation of the superconducting order parameter does not impact on the character of the Edelstein effect. In the performed analysis the other parameters are:  $\alpha=0.1 t$,  $\delta_z=0.5t$, $t'=0.4t$, $t_b=0.75t$ (as in the main text) and $g=2.0 t$.
\label{Fig1supmat}}
\end{figure*}

\section{Self-consistent analysis}

In this Section we present the analysis of the Edelstein effect by considering the self-consistent solution of the superconducting order parameter in real space in the presence of inhomogeneous phase gradient.

Before looking at the spatial dependence we first consider the homogeneous problem and show the dependence of the value of the orbital moment $\langle L_y\rangle$ versus the chemical potential by evaluating the gap self-consistently. The result is shown in Fig.~\ref{Fig0sm}(a), where we clearly see that no difference is visible at bare eye on the signal between the uniform choice $\Delta_i=0.01 t$ (the same as the one shown in Fig. 1(d) of the main text) and the values of the gaps obtained self-consistently for $g=0.8 t$, shown Fig.~\ref{Fig0sm}(b). The magnitude of the gap is reasonable for a BCS superconductor and the signal reaches an asymptotic value by reducing the gap, that substantially depends on the details of the Fermi surface.

Concerning the evaluation of the self-consistent solutions we recall that at a given site $i_x$ the orbital dependent superconducting order parameters $\Delta_{a}$ are expressed as $\Delta_{a}(i_x)=\frac{1}{N_x N_y} \sum_{n} g_{a}\,\langle n| c_{i_x,a\uparrow} c_{i_x,a\downarrow} |n \rangle$,  with $a=(yz,xz,xy)$ and an interaction $g_{a}=g$ that is orbital independent. Here, we performed the computation by self-consistently evaluating the trace of the orbital-dependent pairing operator for the spin-singlet channel, $\hat{P}^{a}(i_x) =c_{i_x,a\uparrow} c_{i_x,a\downarrow}$, over all the eigenstates $|n\rangle$ of the Hamiltonian associated to negative energies $E_{n}<0$ at zero temperature (at finite temperature the trace is over all energy configurations weighted by the Fermi function). Since the eigenstates $|n\rangle$ depend on $\Delta_{a}(i_x)$ and the orbital Rashba interaction couples the crystal wave vector with the in-plane orbital moments, the gap equations of the orbital dependent order parameters are coupled between each other. The procedure is iterated by evaluating the order parameters for all the sites in the slab upon reaching the desired accuracy.

The main outcomes are reported in Fig. \ref{Fig1supmat}. In Fig.~\ref{Fig1supmat}(a) we see that the self-consistent gap amplitudes develop a modulation at short (Fermi) wave lengths in proximity of the phase defect. At the same time, the profile of the in-plane orbital moment $\langle L_y\rangle$, shown in Fig.~\ref{Fig1supmat}(c), is smooth far away from the defect, varies on the length scale of the superconducting coherence length and changes sign in the same fashion as in the case of a uniform profile of the amplitude. Similar behavior but with different spatial distribution is observed in Fig.~\ref{Fig1supmat}(d)-(f) and in Fig.~\ref{Fig1supmat}(g)-(i) for different values of the chemical potential. We point out that the main fingerprints of the Edelstein effect are substantially robust in regimes of electron filling that apply to low density superconductors, as for instance LAO-STO oxide interface, and high density cases as for conventional metallic superconductors.

We thus confirm within the self-consistent approach the robustness of the main fingerprints reported in the main part of the work, that emerged already by assuming a uniform profile of the superconducting order parameters. In particular, we confirm: i) the real space sign change, ii) the spatial variation on the scale of the superconducting coherence length and iii) the large amplitude of the induced orbital moment, that confers a colossal character to the effect. 

\begin{figure*}[tb]
\includegraphics[width=0.95\textwidth]{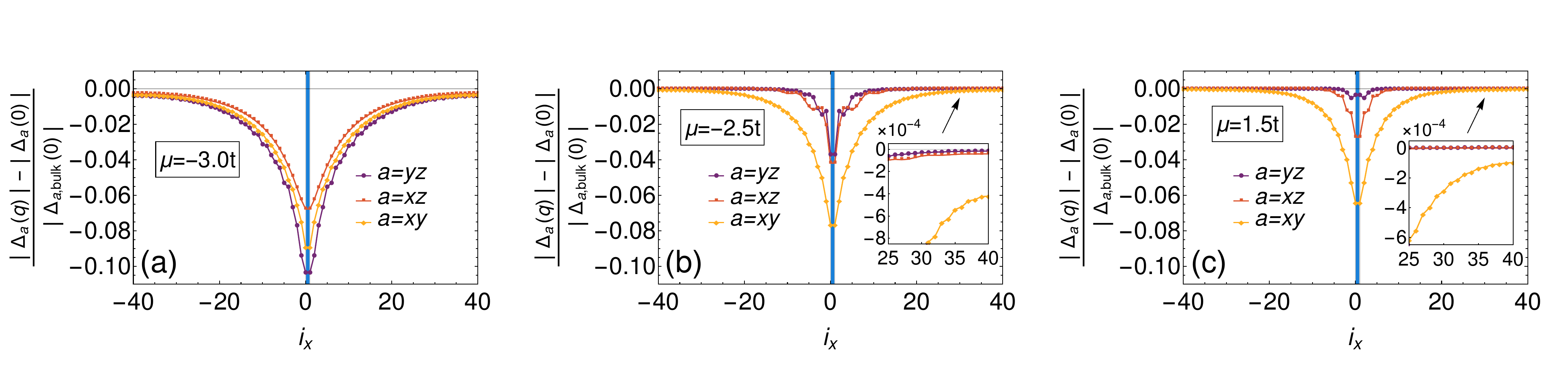}
\caption{Relative difference of the spatial dependent order parameter with non-vanishing phase gradient (i.e. $q=0.01$) as compared to the configuration without a supercurrent bias (i.e. $q=0$). The phase defect is nearby $i_x=0$. We consider three representative filling configurations: (a) $\mu=-3.0 t$, (b) $\mu=-2.5 t$, and $\mu=1.5 t$ for $\alpha_{\text{OR}}=0.1 t$. The variation of the superconducting order parameter is in the range $1\% - 10 \%$. 
\label{Fig2supmat}}
\end{figure*}

\begin{figure}[hb]
\includegraphics[width=0.99\columnwidth]{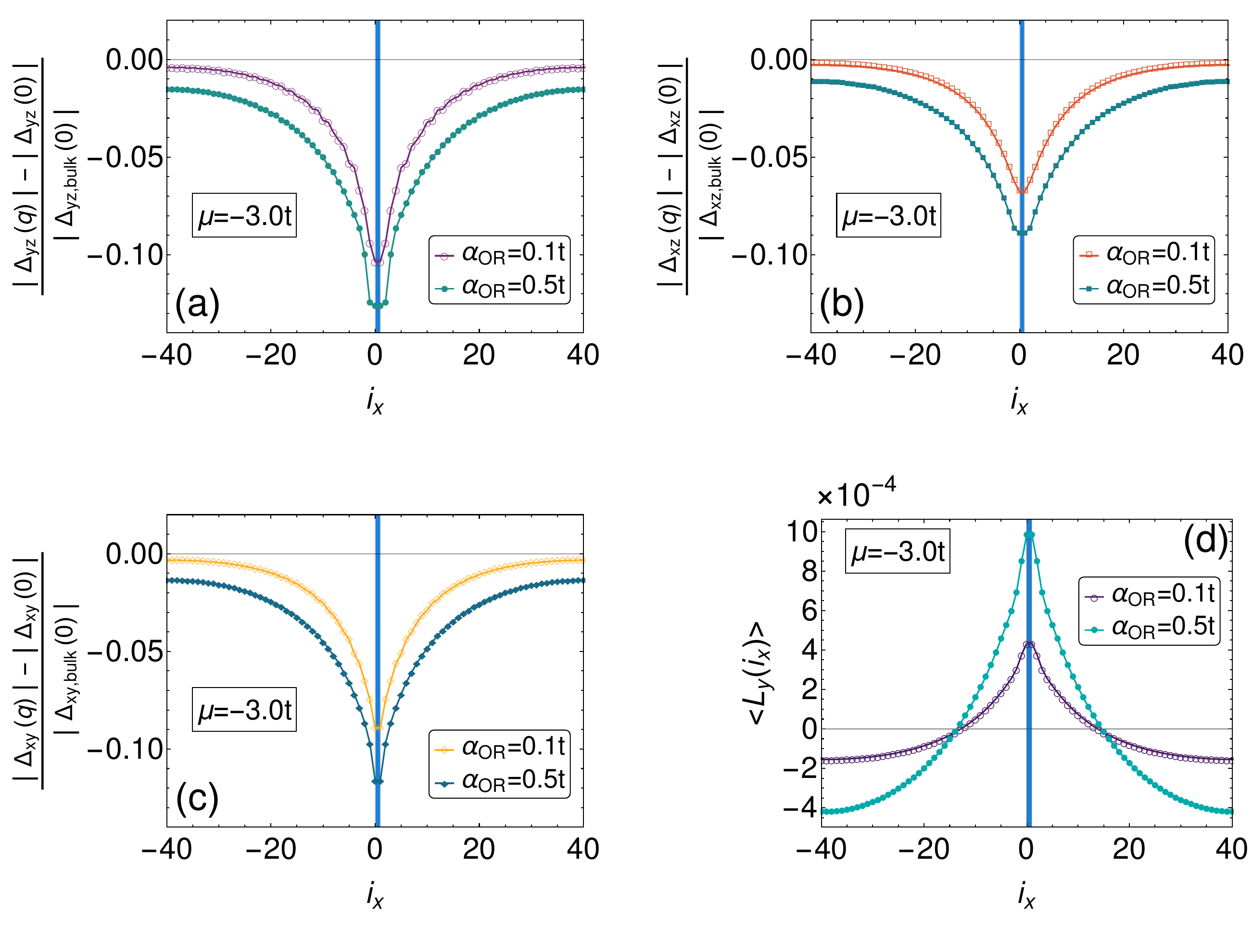}
\caption{Relative difference of the spatial dependent order parameters with non-vanishing phase gradient (i.e. $q=0.01$) as compared to the configuration without a supercurrent bias (i.e. $q=0$) for two different amplitudes of the orbital Rashba coupling, $\alpha_{\text{OR}}=0.1 t$ and $0.5 t$ at $\mu=-3.0 t$. The phase defect is nearby $i_x=0$. Due to the reduced symmetry and the orbital anisotropy, the change of the order parameter is orbital dependent as shown in (a), (b) and (c) for the $yz$,$xz$, and $xy$ order parmaters, respectively. The variation associated with the change of the orbital Rashba coupling is of the order of $2\%$. In (d) we report the induced orbital moments for $\alpha_{\text{OR}}=0.1 t$ and $0.5 t$. 
\label{Fig3supmat}}
\end{figure}

Finally, in order to single out the impact of the orbital moment induced by the supercurrent on the order parameter we consider the relative difference of the spatial dependent order parameters with non-vanishing phase gradient (i.e. $q=0.01$) as compared to the configuration without a supercurrent bias (i.e. $q=0$). In Fig. \ref{Fig2supmat} we firstly show the evolution for different electron densities at a given amplitude of the orbital Rashba coupling.

Furthermore, it is useful to compare the variations for two different amplitudes of the orbital Rashba coupling as shown in Fig. \ref{Fig3supmat}. Since the intensity of the Edelstein effect is proportional to the orbital Rashba coupling, we demonstrate that a change in the orbital moment amplitude (Fig. \ref{Fig3supmat}(d)) leads to a variation of the strength of the superconducting order parameters of few percents. As expected, due to the reduced symmetry induced by the application of the supercurrent and the anisotropy of the multi-orbital model, the modification of the order parameters turns out to be orbital dependent ((Fig. \ref{Fig3supmat}(a)-(c)).

